\documentclass[12pt]{iopart}

\usepackage{epsfig}
\usepackage{graphics}
\usepackage{floatflt}
\usepackage{caption}

\begin{document}

\vspace*{-2cm}
\title[ALICE Electrons from heavy-flavour decays]{Investigation of charm 
and beauty production via semileptonic 
decays of heavy-flavour hadrons in pp at 7 TeV and Pb--Pb at 2.76 TeV with 
ALICE}

\author{S. Masciocchi for the ALICE Collaboration}

\address{GSI Helmholtzzentrum f\"{u}r Schwerionenforschung GmbH,\\
Planckstrasse 1, 64291 Darmstadt, Germany
}
\ead{s.masciocchi@gsi.de}

\begin{abstract}
Electron spectra measured with ALICE at mid-rapidity are used to study the
production of hadrons carrying a charm or a beauty quark.
The production cross section of electrons from heavy-flavour hadron decays is 
measured in pp collisions at $\sqrt{s}$=7~TeV. Electrons from the beauty decays
are identified via the displacement from the interaction vertex.
From the electron spectra measured in Pb--Pb collisions, we determine the nuclear 
modification factor, which is sensitive to the heavy-quark energy loss in a hot
strongly interacting medium.
\end{abstract}


The measurement of heavy-flavour production cross sections allows an important test
for perturbative QCD descriptions of hard processes in hadronic interactions. Moreover,
charm and beauty quarks are a sensitive tool to study the flavour
dependence of parton energy loss in high energy nucleus-nucleus collisions,
the Quark-Gluon Plasma (QGP).
Precision measurements of heavy-flavour production in proton-proton collisions
provide the necessary reference  for the interpretation of the Pb--Pb collision
results within a QGP framework.

One channel to measure the heavy-flavour production cross section
is semi-leptonic decays of hadrons
carrying a charm or a beauty quark. Such decays have a relatively large branching ratio (of about
10\%) and offer a complementary approach to that of the reconstruction in
exclusive hadronic decays.

ALICE (A Large Ion Collider Experiment) \cite{ALICE} at the LHC is well equipped
to address these physics topics: it has
very good particle identification and spatial resolution for the separation of secondary
vertices. The latest ALICE heavy-flavour physics results are summarised in \cite{Andrea}. This
contribution presents the results from the measurement of electron spectra in proton-proton and 
Pb--Pb collisions at mid-rapidity.

\vspace*{-4mm} 
\section{Electron identification and inclusive spectra}
\vspace*{-4mm} 
Electrons are identified at mid-rapidity using the information provided by the Time Projection 
Chamber (TPC), the Transition Radiation Detector (TRD), and the Time-Of-Flight (TOF) detector.
The time-of-flight of the individual particles is required to be consistent with the electron hypothesis.
Using the TRD information, an electron likelihood is computed from the energy deposited in the
6 detector layers by charged particles.
In order to provide a good e/$\pi$ separation, 
tracks are required to have at least 5 out of maximum 6 chambers providing information 
on the energy deposit. A cut providing an 
electron selection efficiency of 80\% is applied to the likelihood value.
Finally, we consider the TPC energy loss d$E$/d$x$ expressed as deviation from the Bethe-Bloch electron curve
and select electron candidates from the top half of the d$E$/d$x$
distribution, to guarantee very high electron purity.

Inclusive electron spectra are based on high-quality tracks passing the particle identification
criteria over a momentum range where the hadron contamination does not exceed 10\%.
In the pp analysis all three detectors were used to identify electrons and the 
spectra extend up to 10 GeV/$c$ in transverse momentum $p_{\rm t}$. In the Pb--Pb analysis 
only the TOF and TPC information were used 
and the spectra extend up to 6 GeV/$c$.
The remaining hadron contamination is 
estimated by fitting the electron and the hadron TPC d$E$/d$x$ Gaussian distributions in 
narrow momentum slices and counting the hadrons passing the PID cut. This contribution is subtracted.

The electron yield 
is corrected for acceptance and for the efficiency of the selection criteria, using Monte-Carlo
simulations. 
An unfolding procedure
is applied to correct for the distortion of the $p_{\rm t}$ distribution due to Bremsstrahlung.

The inclusive electron spectrum contains contributions from many sources other than heavy-flavour 
hadron decays. The most important are: Dalitz decays of light-flavour neutral mesons
($\pi^0$, $\eta$, $\omega$, $\eta$', $\phi$), photon conversions in
the beam pipe and in the detector material, di-electron decays of vector mesons ($\rho$, $\omega$, $\phi$), 
heavy quarkonia, real and virtual QCD photons.
The spectrum composed of these background electrons is described by the so-called electron
cocktail,
 calculated on the basis of meson spectra measured with the ALICE detector,
as described in
more detail elsewhere \cite{SilviaHP}. Further components are parametrized according to other measurements
at the LHC (e.g. quarkonia) or theoretical predictions (e.g. QCD photons).
In this analysis, 
the statistical subtraction of the cocktail
from the inclusive electron spectrum
provides the transverse momentum distribution of electrons from
heavy-flavour hadron decays.

\vspace*{-4mm} 
\section{Results from proton-proton collisions at $\sqrt{s}$=7 TeV}
\begin{figure}[t]
\begin{center}
\includegraphics[width=.45\textwidth]{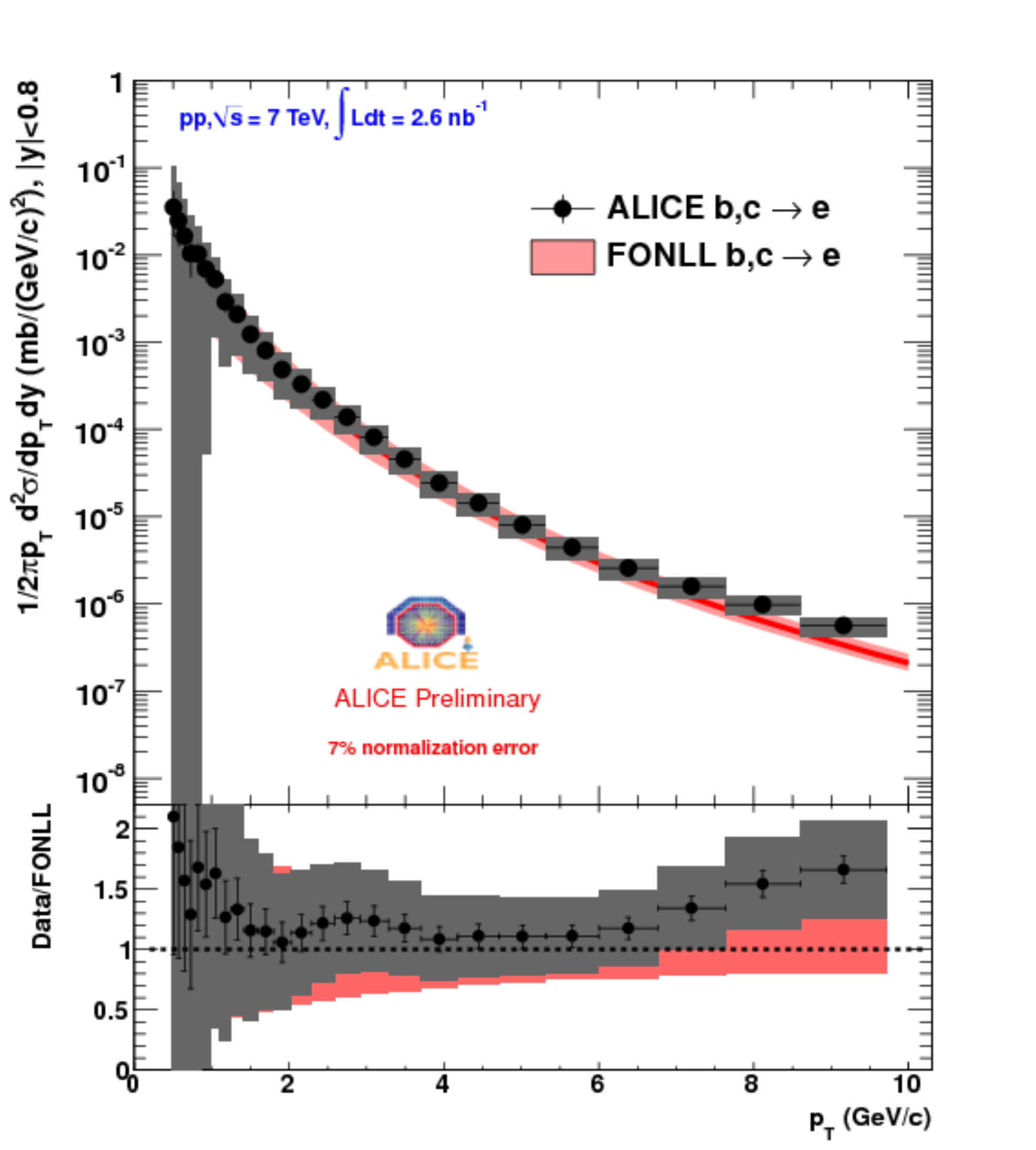}
\includegraphics[width=.45\textwidth]{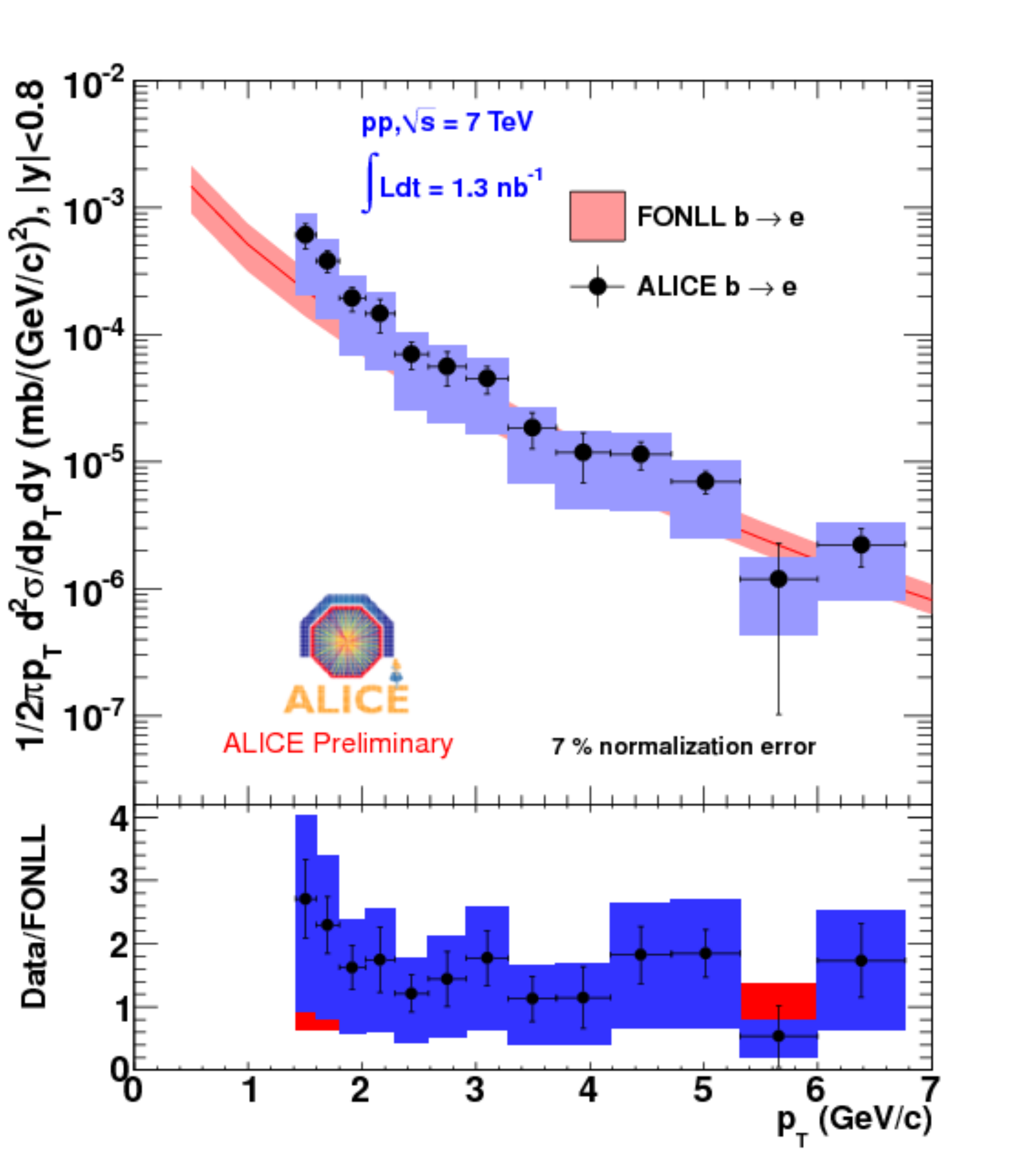}
\captionsetup{singlelinecheck=off,justification=RaggedRight}
\caption{Production cross section of electrons from the decay of hadrons with charm or
beauty (on the left) or only beauty (on the right) content, compared to FONLL predictions (see text),
in proton-proton collisions at $\sqrt{s}$=7~TeV.}
\label{Fig1}
\end{center}
\end{figure}

\vspace*{-4mm} 
A sample of minimum bias pp collisions 
(2.6 nb$^{-1}$) 
recorded by ALICE in summer 2010 at $\sqrt{s}$=7~TeV
was used to measure the production cross section of electrons from heavy-flavour decays
according to the method discussed above. The measured spectrum, including both the charm
and the beauty contributions, is shown in Fig.~\ref{Fig1} on the left. The 
present conservative
estimate of the systematic uncertainty
is $\approx$20\% on the measured spectrum and 
$\approx$10\% on the electron cocktail.

The measured cross section is compared to a prediction at Fixed-Order plus Next-to-Leading Logarithms (FONLL)  
\cite{Matteo} which provides a good description.

The silicon Inner Tracking System (ITS) provides excellent impact parameter resolution
allowing for isolation of decay products of longer living particles. Electrons from the decay of hadrons with 
beauty content are selected by requiring large displacement from the interaction
vertex.
The impact 
parameter resolution is well described in our simulations.
A selection is applied by requiring the electron to miss the interaction vertex by at least
three times the error on the separation of the track from the vertex (of the order of 80~$\mu$m
at 2 GeV/$c$).
The remaining contribution from charm decays is estimated from the D meson cross section
at mid-rapidity, and subtracted.
The resulting spectrum of
electrons from beauty decays is shown in Fig.\ref{Fig1} on the right, also
compared to the FONLL prediction for the beauty component.
This is the first 
beauty production cross section measurement in ALICE.

\vspace*{-4mm} 
\section{Results from Pb--Pb collisions at $\sqrt{s_{\rm NN}}$=2.76 TeV}
\vspace*{-4mm} 
A similar analysis is performed with Pb--Pb collisions recorded by ALICE in November and 
December 2010.
Electrons are identified with the TOF and TPC detectors and inclusive
electron spectra are obtained in 6 different centrality bins.
The hadron contamination in the electron sample remains below 10\% in the momentum range
between 1.5 and 6 GeV/$c$, to which this first measurement is restricted.

The inclusive electron spectra are corrected for acceptance and efficiencies using a HIJING
Monte-Carlo sample enriched with heavy-flavour signals. An electron cocktail for each centrality bin
is calculated based on the charged pion spectra measured by ALICE.
In the Pb--Pb analysis the systematic uncertainties currently are large: of the order of 
35\% on the inclusive spectra and 25\% on the cocktails. 

When comparing the inclusive electron spectra with those generated using the
cocktail prescription (see Fig. 2 and 3 in \cite{Yvonne}, where more details are
discussed), a hint for an
excess is observed in the electron yield in the transverse momentum region between 1.5 and $\approx$3.5 GeV/$c$.
The excess (indicated by the ratio in Fig. 3 in \cite{Yvonne}, significantly larger than in pp collisions, despite the large
systematic uncertainties) 
exhibits a clear centrality dependence. 
Cross checks
with the ALICE measurement of D mesons at mid-rapidity seem to exclude an additional
charm source. 
The possibility of a 
contribution from thermal radiation, already measured 
at RHIC \cite{TheRad}, will be further investigated.

We compute the nuclear 
modification factor $R_{\rm AA}$ defined as
$R_{\rm AA}(p_{\rm t}) = \frac{1}{\langle T_{\rm AA}\rangle} \cdot \frac{{\rm d}N_{\rm AA}/{\rm d}p_{\rm t}}{{\rm d}\sigma_{\rm pp}/{\rm d}p_{\rm t}}$, 
where  \mbox{$\langle T_{\rm AA}\rangle$} is the average nuclear overlap function for a given centrality bin,
and ${\rm d}N_{\rm AA}/{\rm d}p_{\rm t}$ and ${\rm d}\sigma_{\rm pp}/{\rm d}p_{\rm t}$ describe the electron yield in Pb--Pb
and in pp collisions. 

\begin{floatingfigure}[l]{0.5\linewidth} 
\hspace*{-12mm}
\includegraphics[width=.5\textwidth]{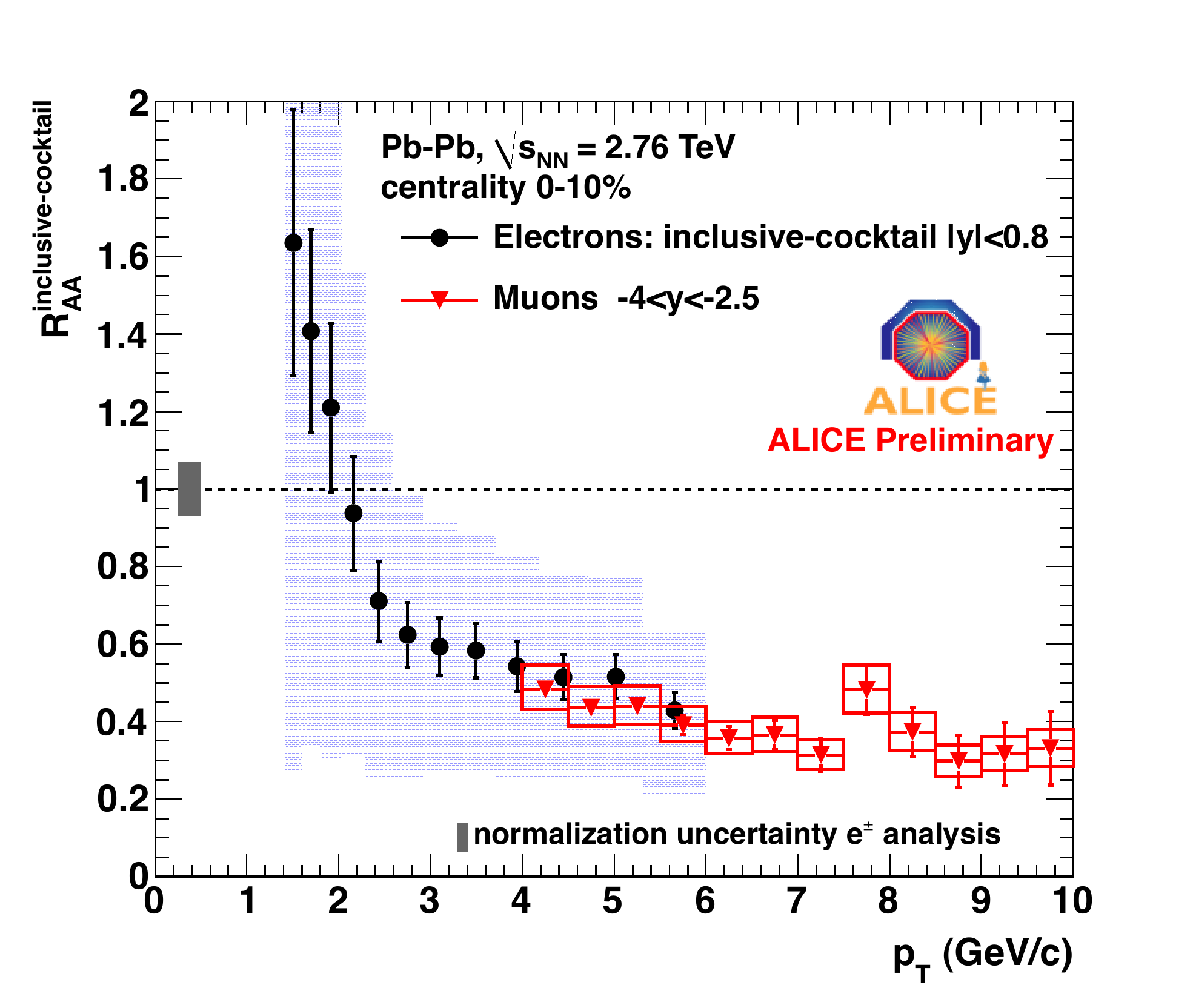}
\captionsetup{singlelinecheck=off,justification=RaggedRight}
\caption{Nuclear modification factor for single electrons (inclusive - cocktail) at mid-rapidity and
single muons at forward rapidity.}
\label{Fig2}
\end{floatingfigure}

The latter is obtained by extrapolating the
measured spectrum at $\sqrt{s}$=7~TeV down to the energy of 2.76~TeV 
according to FONLL prescriptions \cite{Extrap}.

The resulting $R_{\rm AA}$ for the cocktail subtracted electrons in the 0-10\% most
central events is shown in Fig.~\ref{Fig2}. In the transverse momentum region between 3.5 and
6 GeV/$c$, in which we assume that the cocktail-subtracted electron spectra are dominated
by electrons from heavy-flavour decays, 
a suppression
by a factor 1.5-4 is observed.
This is similar to the suppression measured with single muons 
at forward rapidity \cite{Xiaoming}.
This measurement is compatible with in-medium energy loss of the heavy-flavour quarks.

\vspace*{-2mm} 
\section{Conclusions}
\vspace*{-4mm} 
The measurement of electron spectra at mid-rapidity in ALICE allows to determine
the production cross section of electrons from heavy-flavour decays in pp collisions
at $\sqrt{s}$=7~TeV. The first measurement
of beauty hadron decays was presented. FONLL predictions provide a good
description of the measured spectra.
Together with the electron spectra measured in Pb--Pb collisions, a nuclear modification
factor is extracted, showing
a strong suppression of electrons from heavy-flavour decays in
central collisions.

The next goal is to measure the suppression factor for electrons from 
charm and beauty hadrons separately, to provide constraints to in-medium
energy loss models.

\end{document}